\documentstyle[12pt]{article}
\begin{document}

\expandafter\ifx\csname amssym.def\endcsname\relax \else\endinput\fi
\expandafter\edef\csname amssym.def\endcsname{%
       \catcode`\noexpand\@=\the\catcode`\@\space}
\catcode`\@=11

\def\undefine#1{\let#1\undefined}
\def\newsymbol#1#2#3#4#5{\let\next@\relax
 \ifnum#2=\@ne\let\next@\msafam@\else
 \ifnum#2=\tw@\let\next@\msbfam@\fi\fi
 \mathchardef#1="#3\next@#4#5}
\def\mathhexbox@#1#2#3{\relax
 \ifmmode\mathpalette{}{\m@th\mathchar"#1#2#3}%
 \else\leavevmode\hbox{$\m@th\mathchar"#1#2#3$}\fi}
\def\hexnumber@#1{\ifcase#1 0\or 1\or 2\or 3\or 4\or 5\or 6\or 7\or 8\or
 9\or A\or B\or C\or D\or E\or F\fi}

\font\tenmsa=msam10
\font\sevenmsa=msam7
\font\fivemsa=msam5
\newfam\msafam
\textfont\msafam=\tenmsa
\scriptfont\msafam=\sevenmsa
\scriptscriptfont\msafam=\fivemsa
\edef\msafam@{\hexnumber@\msafam}
\mathchardef\dabar@"0\msafam@39
\def\dashrightarrow{\mathrel{\dabar@\dabar@\mathchar"0\msafam@4B}}
\def\dashleftarrow{\mathrel{\mathchar"0\msafam@4C\dabar@\dabar@}}
\let\dasharrow\dashrightarrow
\def\ulcorner{\delimiter"4\msafam@70\msafam@70 }
\def\urcorner{\delimiter"5\msafam@71\msafam@71 }
\def\llcorner{\delimiter"4\msafam@78\msafam@78 }
\def\lrcorner{\delimiter"5\msafam@79\msafam@79 }
\def\yen{{\mathhexbox@\msafam@55 }}
\def\checkmark{{\mathhexbox@\msafam@58 }}
\def\circledR{{\mathhexbox@\msafam@72 }}
\def\maltese{{\mathhexbox@\msafam@7A }}

\font\tenmsb=msbm10
\font\sevenmsb=msbm7
\font\fivemsb=msbm5
\newfam\msbfam
\textfont\msbfam=\tenmsb
\scriptfont\msbfam=\sevenmsb
\scriptscriptfont\msbfam=\fivemsb
\edef\msbfam@{\hexnumber@\msbfam}
\def\Bbb#1{{\fam\msbfam\relax#1}}
\def\widehat#1{\setbox\z@\hbox{$\m@th#1$}%
 \ifdim\wd\z@>\tw@ em\mathaccent"0\msbfam@5B{#1}%
 \else\mathaccent"0362{#1}\fi}
\def\widetilde#1{\setbox\z@\hbox{$\m@th#1$}%
 \ifdim\wd\z@>\tw@ em\mathaccent"0\msbfam@5D{#1}%
 \else\mathaccent"0365{#1}\fi}
\font\teneufm=eufm10 scaled 1200
\font\seveneufm=eufm7
\font\fiveeufm=eufm5
\newfam\eufmfam
\textfont\eufmfam=\teneufm
\scriptfont\eufmfam=\seveneufm
\scriptscriptfont\eufmfam=\fiveeufm
\def\frak#1{{\fam\eufmfam\relax#1}}
\let\goth\frak

\csname amssym.def\endcsname
\input amssym.def
\def\trait{\hrule width5cm height1pt \vskip1pt \hrule width6cm}
\def\psaut{\vskip 5pt plus 1pt minus 1pt}
\def\saut{\vskip 10pt plus 2pt minus 3pt}
\def\gsaut{\vskip 20pt plus 3pt minus 4pt}
\def\square{\hfill\hbox{\vrule height .9ex width .8ex
  depth -.1ex}}
\def\sqr#1#2{{\vcenter{\vbox{\hrule height.#2pt
\hbox{\vrule width.#2pt height#1pt \kern #1pt
\vrule width.#2pt}
\hrule height.#2pt}}}}
\def\bsquare{\sqr34}
\def\fl{\rightarrow}
\def\flq#1{\buildrel {#1} \over \longrightarrow}
\def\ds{\displaystyle}
\gsaut
\gsaut
\gsaut
\gsaut
\gsaut
\gsaut
\gsaut
\gsaut

\centerline{\bf{A STAR-PRODUCT APPROACH TO NONCOMPACT QUANTUM GROUPS.}}
\saut
\centerline{ by Fr\'ed\'eric BIDEGAIN and Georges PINCZON}
\gsaut
\small
\centerline{Laboratoire d'Alg\`ebre et Analyse:
Th\'eorie des repr\'esentations}
\centerline{Physique Math\'ematique, Universit\'e de Bourgogne}
\centerline{B.P. 138, F-21004 DIJON Cedex, France.}
\normalsize
\gsaut
\gsaut
 {\bf{Abstract:}}
\psaut
Using duality and topological theory of well behaved Hopf algebras (as defined
in [2]) we construct star-product models of non compact quantum groups from
Drinfeld and Reshetikhin standard deformations of enveloping Hopf algebras
of simple Lie algebras. Our star-products act not only on coefficient
functions of finite-dimensional representations, but actually on all
$C^\infty$ functions, and they exist even for non linear (semi-simple)
Lie groups.
\gsaut
\gsaut
\gsaut
\gsaut
\small
AMS classification : primary 17B37, 16W30, 22D05, 46H99, 81R50.
\psaut
\centerline{Running title : noncompact quantum groups.}
\footnotesize
\psaut
\centerline{[ To be published in ``Letters in Mathematical Physics" ]}
\normalsize

\vfill\eject

{\bf Introduction.}
 \saut
In the star-product approach to quantum mechanics [1], the observables are
usual functions on phase space, so the geometry is not changed, but the product
of observables is no longer the commutative one; the new product (star-product)
is a deformation [6] of the initial one, with parameter $\hbar$ and
leading cocycle the Poisson bracket. The simplest example is the Moyal product,
which is completely explicit. By analogy, in his fundamental Berkeley paper on
quantum groups [3], Drinfeld insisted on the similitude which should exist
between star-product formulation of quantum mechanics, and ``functional
realizations'' (cf. e.g.: [5] or [15]) of quantum groups: these
realizations should be deformations of the commutative product of functions on
the corresponding group, with (skew symmetrized) leading cocycle a Poisson
bracket. In [2], this nice interpretation was given a precise mathematical
justification (using adapted duality arguments and triviality results of
Drinfeld [4]) in the case of compact Lie groups : it was shown that all
standard ``functional'' models (e.g.: [5]) can be realized in such a
way. Let us go further in the discussion: in quantum mechanics, the
star-product is fortunately not limited to act on polynomial functions, but
actually lives on $C^\infty$-functions. On the other hand, all known
``functional'' models of quantum groups are given by generators and relations,
so when realized as preferred deformations [7], as in [2], they live on
coefficient functions of finite dimensional (f.d.) representations. This
restriction seems unnatural, and only justified by necessity : FRT-models were
initially defined by generators and relations! Actually, and this is one of the
main results of [2], the models do extend to all $C^\infty$-functions, as
they should, and therefore the similitude with the star-product approach to
quantum mechanics is complete. This being noted, the coefficient functions are
a sub-Hopf algebra for the star-product, which can be seen as a dense core on
which the star-product can be more easily computed.

In the present paper we want to show how the results of [2] for
compact Lie groups can be extended to connected semi-simple Lie groups. This
will answer various attempts to define ``non-compact'' quantum
groups, and also give a serious framework for a discussion of real forms (see
e.g.: [13]). To be precise, if one wants to construct a star-product only
on the space of f.d. representations coefficients, then it is a straightforward
improvement of [2], because we show that the algebra of coefficients of f.d.
representations of a connected semi-simple Lie group can always be identified
with the same algebra for a suitable compact Lie group (an avatar of the
Weyl unitary trick; as a consequence this algebra is a finitely generated
domain). By the way, we show that the core of coefficient functions is dense in
$C^\infty$-functions if and only if the group is linear, a result which
stresses the difference between the linear and non linear cases.

As mentionned above, in our opinion, the coefficient space is much too
restrictive for star-product theory, and the good space to deal with is
the space of $C^\infty$-functions. So we show that in every case (linear
or not) the  star-product extends to all $C^\infty$-functions: starting
with a Hopf deformation of the enveloping algebra of a semi-simple complex
Lie algebra $\goth g$ (such as Drinfeld models [3], or Reshetikhin models
[11]), we construct a preferred deformation [7] of the algebra of
$C^\infty$-functions on any Lie group with Lie algebra a real form of
$\goth g$. This is completely new for semi-simple Lie groups which
are not linear, a case in which it is obviously meaningless to try
and define quantum models by standard FRT-techniques of generators and
relations. We think that this result will lead to
new models defined by generators and relations, but acting on coefficients of
infinite dimensional representations (such as the discrete series when it
exists), and we intend to develop this aspect in a subsequent paper.
\gsaut
{\bf{1. Hopf algebras of functions and distributions associated
 to a connected Lie group}}
  \saut
Let $G$ be a connected Lie group, and $H(G) = C^\infty (G)$. When endowed
with its usual Fr\'echet topology, $H(G)$ is a nuclear Montel space;
it is well known that $H(G\times G) = H(G) \hat\otimes H(G)$.
Now, the commutative product on $H(G)$ defines a topological algebra structure.
Moreover, setting $\delta	 (f) (x, y) = f(xy), f \in H(G), x, y \in G$, we get
a continuous coproduct. There is a unit (the constant 1 function), a counit
(the Dirac distribution $\partial_e$), and an antipode defined by
$S_0 (f) (x) = f(x^{-1}), f \in H(G), x\in G$. Since all the mappings involved
 are continuous, $H(G)$ is a well behaved topological Hopf algebra ([2] (1.2)).
\saut

Now let $A(G) = H(G)^*$ be the space of compactly supported distributions
on $G$, with strong dual topology; by ([2] (1.3)), the transposition defines
a well behaved topological Hopf algebra structure on $A(G)$: product is
the convolution product, the unit is the Dirac distribution $\partial_e$
the counit is the evaluation on 1. In order to check the coproduct, we
introduce
$\partial: G\rightarrow A(G)$ defined by $< \partial_x \mid f > = f (x),
f \in H(G), x\in G$; it is easily seen that $\partial$ actually defines a
topological inclusion of $G$ as a closed subset of $A$, so in the sequel
we identify $G$ and $\partial(G)$. This being done, one has $G^\bot=\{0\},$ so
$\overline{\hbox{Vect}(G)} = A(G)$, and the coproduct $\triangle_0 :
A(G) \rightarrow A(G) \hat\otimes A(G) = A(G\times G)$ is given on
$G$ by $\triangle_0 (x) = x\otimes  x, x\in G$. \saut

Let ${\goth g}_0$ be the Lie algebra of $G$, $\goth g$ its complexification,
and ${\cal U} ({\goth g})$ the corresponding enveloping algebra.
We identify, as usual, ${\cal U}({\goth g})$ and the algebra of left invariant
differential operators of finite order on $G$; the left regular representation
of $G$ on $H(G)$ defined by $L_g (f) (x) = f(g^{-1} x),\hskip .1cm f\in
H(G), g, x \in G, \hbox{ is a } C^\infty$ representation of $G$ [14] and the
same holds for its contragredient [14] :
$$<\breve{L}_g (a) \mid f > = < a\mid L_{g^{-1}}(f) >, a \in A, f \in H,
g\in G$$.
But $\breve{L}_g (a) = \partial_g . a, \hbox{ so taking } a= \partial_e\hbox{
we see that } \partial: G\rightarrow A \hskip.1cm\hbox{is a } C^\infty$
mapping; we define a linear map $ i: {\cal U} ({\goth g})\rightarrow A$
by $i(u)(f) = \partial_e[u(f)], u  \in {\cal U}(g), f\in H(G)$.
By left invariance one has $u(f)_g = i(u) [L_{g^{-1}}
(f)]$, so $i$ is one to one, and it is easy to check that $i$ is a morphism,
so we  identify $\cal U $ and $i({\cal U})$ and consider in the sequel that
${\cal U}(g) \subset A$. Since
$$<\frac{d}{dt} (\hbox{exp} t X) _{t=0} \mid f > = <\frac{d}{dt}
(\breve{L}_{\hbox{exp} t\hskip.1cm X} (\partial_e))_{t=0} \mid f >$$
$$= \frac{d}{dt} <\partial_e \mid L_{exp\hskip.1cm (-tX)} (f) >_{t=0} =
X(f)_e = i (X) (f)$$
one has   $\frac{d}{dt} (\hbox{exp} t\hskip.1cm X)_{t=0} = X$, in $A(G)$.
Therefore,  by differentiation of
$$\triangle_0 (\hbox{exp}t\hskip.1cm X) = \hbox{exp}t X \otimes
\hbox{exp}t\hskip.1cm X,$$
we deduce that $\triangle_0(X) = X \otimes 1 + 1 \otimes X,
X\in {\goth g}_0$. By the same technique, the
antipode of $A(G)$ restricts on ${\cal U}({\goth g})$ to the usual antipode
of ${\cal U} ({\goth g})$, and finally ${\cal U} ({\goth g})$ is a Hopf
subalgebra of $A(G)$. Another interpretation of this result is as follows:
\hskip.1cm $ {\cal U}({\goth g}) $ has a natural Hopf structure, which extends
to a Hopf structure on $A(G)$, though ${\cal U}({\goth g})$ is
certainly not dense in $A(G)$. This remark is the key to understand the
star-product realization of quantum groups that we shall give later.
 \gsaut
{\bf{2. Hopf algebras of coefficients and generalized distributions associated
to a connected semi-simple Lie group}}
\saut

Let $G$ be a connected semi-simple Lie group. Given a finite dimensional (f.d.)
representation $\pi$ of $G$, and any $M \in {\cal L}(V_\pi)$, the coefficient
of $\pi$ associated to $M$ is defined by $C_M^\pi (x) = \hbox{Tr}(M\pi(x)),
x\in G$, and is an analytic function on $G$. When $\pi$ is irreducible, by
Burnside theorem, $C^\pi: {\cal L}(V_\pi)\rightarrow H(G)$ is one to one,
and we define
$H(G)_\pi = C^\pi ({\cal L}(V_\pi)).$ Then we fix once for all a set $\Pi$ of
f.d. irreducible representations of $G$ which contains one and only one element
of each equivalence class, and such that if $\pi \in \Pi$ and
$\pi \neq \breve{\pi}$, then ${\displaystyle \breve{\pi} \in \Pi} $.
\psaut

Let ${\cal H}(G)$ be the subspace of $G$ generated by the coefficients of all
f.d.  representations.
\saut

{\bf(2.1) {Lemma}}:\hskip.1cm ${\cal H} (G) ={\displaystyle
 \bigoplus_{\pi \in \Pi} }\hskip.1cm H(G)_\pi$
\saut

{\bf{Proof}}: Since $\goth g$ is semi-simple, any representation of $G$ is
 semi-simple, so  ${\cal H}(G) = {\displaystyle \Sigma_{\pi \in \Pi}}
H(G)_\pi$.
Let ${\goth g}_0 = {\goth k} + {\goth p}$ be a Cartan decomposition of
${\goth g}_0$, $\goth u$ the associated compact real form of $\goth g$ ,
and $U$ the compact simply connected group with Lie algebra $\goth u$.
Given $\pi \in \Pi$, $d\pi $ is a representation of $\goth g$, therefore
of $\goth u$, so $\pi$ is actually a representation of $U$, and $C_M^\pi$
is also an analytic function on $G$. By the Peter-Weyl theorem
$\displaystyle \Sigma_{\pi \in \Pi} H(U)_\pi$ is a direct sum in
$H(U)$; assume now that $\Sigma_\pi\hskip.1cm f_\pi =0, f_\pi =
C_{M_\pi}^\pi\hskip.1cm \in H(G)_\pi$. Using Taylor formula, we deduce that
$\Sigma_\pi\hskip.1cm f_\pi (X^n) = 0,
\forall \hskip.1cm X \in {\goth g}_0, \forall  n \geq 0$, so
$\Sigma_\pi \hskip.1cm f_\pi =0$ on ${\cal U}({\goth g})$. Using once more
Taylor formula and the analyticity of $\Sigma_\pi \hskip.1cm f_\pi \hbox{ on }
U$, $\Sigma_\pi \hskip.1cm f_\pi = 0 \hbox{ on } U, \hbox{so }\hskip.1cm
f_\pi = 0 \hbox{ on}\hskip.1cm U, \forall \pi$, therefore $M_\pi = 0,
\forall \pi$, and then $f_\pi = 0 \hskip.1cm\hbox{on } G, \forall \pi$.
$\square$
\psaut

Now ${\cal H}(G)$ is a Hopf subalgebra of $H(G)$, that we shall call the
{\it{Hopf algebra of coefficients of $G$}}. This is seen as follows:
 As in ([2] (2.3)), one has:\hskip.3cm $C_M^\pi \otimes C_{M'}^{\pi'} =
 C_{M \otimes M'}^{\pi \otimes \pi'}$, if $\pi \otimes \pi'$ denotes the
 usual tensor product of the representations $\pi$ and $\pi'$ of $G$,
 so ${\cal H}(G)$ is a subalgebra. Given a basis $\{e_i\}$ of
$V_\pi$, let $M=e_i^* \otimes e_j$, then $C_M^\pi (x) = \pi_{ij}(x),
\Delta (\pi_{ij}) ={\displaystyle \Sigma_k} \pi_{ik} \otimes \pi_{kj}$, so
$\delta({\cal H}(G)) \subset {\cal H}(G) \otimes {\cal H}(G)$. It is clear
that the  unit of $H(G)$ is in ${\cal H}(G)$, and moreover the counit of
$H(G)$ defines a counit on ${\cal H}(G)$. Finally, $S_0 (	\pi_{ij}) =
\breve{\pi}_{ji}$, if $\displaystyle{\breve{\pi}}$ denotes the contragredient
of $\pi$, so $S_0({\cal H}(G)) \subset {\cal H}(G)$. \psaut

Finally, we note that since the elements of ${\cal H}(G)$ are analytic
functions on $G$, ${\cal H}(G)$ {\it{is actually a domain}}. Moreover,
with the notations of the proof of Lemma (2.1), ${\cal H}(G) \subset
{\cal H} (U)$, and it is well known that ${\cal H}(U)$ is of countable
 dimension, so the same holds for ${\cal H}(G)$. This last remark and
 ([2] (1.5.1)) show that, when endowed with its natural topology,
 ${\cal H}(G)$ is a well behaved topological Hopf algebra, so its dual
 ${\cal A}(G) =  {\displaystyle{\prod_{\pi \in \Pi}}} {\cal L}(V_\pi)$
 is also a topological Hopf algebra (with Fr\'echet product topology)
 for the transposed structure, if we define the duality (as in [2])
 by $< \Sigma \hskip.1cmM_\pi \hskip.1cm\mid\hskip.1cm \Sigma\hskip.1cm
 C_{N_{\pi}}^\pi > = \Sigma\hskip.1cm \hbox{Tr} (M_{\pi} \hskip.1cmN_{\pi}
)$; then the product in $A(G)$ is given by: $$\Sigma\hskip.1cm M_\pi
\circ\hskip.1cm \Sigma M'_\pi = \Sigma\hskip.1cm M_\pi\hskip.1cm
\circ\hskip.1cm  M'_\pi,\hskip.2cm M_\pi, M'_\pi \in {\cal L} (V_\pi).$$
Clearly, any f.d. representation of $G$ extends to a representation of
${\cal A}(G)$. To go further, we have to give some details about the
inclusion ${\cal H}(G) \subset H(G)$, which is actually a continuous map.
This is done in the following proposition:
\saut

{\bf{(2.2) Proposition: }} {\it{${\cal H}(G)$ is dense in $H(G)$
 if and only if $G$ has a faithful finite dimensional representation.}}
\saut

{\bf{Remark}} : By a classical result of Harish Chandra ([8], or see the
 proof of \hskip.1cm (2.4)), the condition of (2.2) is equivalent
 to the fact that ${\cal H} (G)$ seperates between points of $G$.
 If this condition is satisfied, by a result of Goto [14], $G$ being
 a semi-simple analytic subgroup of some $GL(p,\Bbb C)$ is actually closed
 in that $GL(p,\Bbb C)$, so we can consider that $G \subset GL(p, \Bbb C)$,
 with the induced topology; in the sequel we shall call such a group a
semi-simple linear group. Note that a semi-simple linear group has finite
center, but the converse is false (e.g.: the metapletic covering of
$SL(2,\Bbb R)$).
\saut

{\bf{Proof of (2.2)}}: If ${\cal H}(G)$ is dense in $H(G)$, apply the Harish
Chandra result. Conversely if $G$ is a closed linear subgroup of $GL(p,\Bbb
C)$,
by a standard partition of unity argument, any $C^\infty$ function on $G$ can
be extended to a $C^\infty$ function on $GL(p,\Bbb C)$. But $GL(p,\Bbb C)$ has
a global chart, namely if $M= (m_{i j}) \in GL(p,\Bbb C), (x_{ij} =
\hbox{Re}(m_{ij}), y_{ij} = \hbox{Im}(m_{ij}))$ defines a global coordinate
system. By a slight and well known improvement
of the Stone-Weierstrass theorem ([12]), on any open set of ${\Bbb R}^{p^2}$,
polynomial functions are dense in $C^\infty$ functions, so (2.2) is true for
$GL(p,\Bbb C)$ (though it is not semi-simple). By restriction to $G$,
polynomials in coefficients of the natural representation of $G$
and of its complex conjugate are dense in $C^\infty (G)$.
$\square$
\saut

So when $G$ is semi-simple and linear, from the inclusion ${\cal H}(G) \subset
H(G)$ and the density of ${\cal H}(G)$ in $H(G)$, we can deduce an inclusion of
$A(G)$ in ${\cal A}(G)$. Since $G$ and ${\cal U} ({\goth g})$ are contained in
$A(G)$, it results, in that case, that $G \subset {\cal A}(G)$ and
${\cal U}({\goth g}) \subset {\cal A}(G)$. Moreover one has
$\overline{\hbox{Vect}(G)} = {\cal A}(G)$. Now it is natural to ask what
happens when $G$ is no longer linear, and the answer is given by:
\saut

{\bf{(2.3) Proposition}}: {\it{One has ${\cal U}({\goth g}) \subset
 {\cal A}(G)$, and $\overline{{\cal U}({\goth g})} = {\cal A}(G)$.}}
\saut

{\bf{Proof}}: Let $G'$ be the adjoint group of $G$, which is a quotient of
 $G$ and has the same Lie algebra. The set $\Pi'$ of irreductible
 f.d. of $G'$ is a subset of the set $\Pi$ of irreductible f.d.
 representations of $G$. So if $u \in {\cal U}({\goth g})$ satisfies
 $\pi(u) =0, \forall \pi \in \Pi$, since $G'$ is semi-simple linear one has
 $u=0$, which proves that ${\cal U}({\goth g}) \subset {\cal A}(G)$.
Finally the algebra ${\cal A}(G)$ is the bicommutant of the semi-simple
representation ${\displaystyle{\oplus_{\pi \in \Pi}}} \pi$ of
${\cal U}({\goth g})$, so density follows from the Jacobson density theorem.
$\square$
\saut
{\bf {(2.4) Remarks}}: (1) When $G$ is not linear one has a map
 $\oplus_{\pi \in \Pi } \pi: G \rightarrow {\cal A}(G)$, with dense range,
 but which is not one to one.
\psaut

(2) When $G$ is linear, the inclusion ${\cal H}(G) \subset H(G)$ with dense
range leads, by transposition ([2] (1.3)), to an inclusion of topological
Hopf algebras $A(G) \subset {\cal A	}(G)$; it results for instance that the
coproduct of $A(G)$ is the restriction of the coproduct of ${\cal A}(G)$,
and so, as proved in (section 1), one has
$\triangle_0 (x) = x \otimes x, x \in G, \hbox{ and}
\hskip.1cm\triangle_0 (X) = X \otimes 1 + 1 \otimes X, X \in {\goth g}$.
So, as in the case of $A(G)$, the Hopf structure of ${\cal A}(G)$ is
the extension of the natural Hopf structure of ${\cal U}({\goth g})$.
{}From the proof of (2.3), this last result holds also if $G$ is not linear.
\psaut
(3) Let us note that $H(G) = C^\infty (G)$ completely specifies $G$, in the
following sense: $G$ is exactly the set of group-like elements of
$A(G) = H(G)^*$, and the topology of $G$ is inherited from the topology
of $A$. On the other hand, when $G$ is not linear, $G$ cannot be contained
in ${\cal A} (G)$ (because of the abovementioned result by Harish Chandra).
Actually ${\cal H}(G) \hbox{ and }{\cal A} (G)$ are related to a compact
group, as shown by:
\saut

{\it{{\bf {(2.5) Proposition}}: There exists a compact connected Lie group
 $K$, with Lie algebra the compact real form of $\goth g$ associated to
${\goth g}_0$, such that ${\cal H}(G) ={\cal H}(K)$.}}
\saut
{\bf {Corollary}}: {\it{${\cal H}(G)$ is a finitely generated algebra.}}
\saut

{\bf{Proof}}: Since ${\cal H}(G \times G') = {\cal H}(G) \otimes {\cal H}(G')$,
 we can assure that $G$ is simple. Let ${\goth g}_0 = {\goth k}+{\goth p}$ be
 a Cartan decomposition of ${\goth g}_0$, ${\goth u} = {\goth k} + i{\goth p}$
the associated compact real form of $\goth g$, and $U$ the simply connected
compact simple group with Lie algebra $\goth u$. Any f.d. representation
$\rho$ of $G$ is actually also a representation of $U$, since $d\rho$ extends
to $\goth g$, and equivalence and irreducibility are preserved. Assume $\rho$
non trivial. Let $K_\rho = \rho(U)$, by the same argument, any f.d.
representation $\theta$ of $K_\rho$ is actually also a representation of
$\tilde{G}$, the universal covering of $G$. Let us show that $\theta$ is
in fact a representation of $G$: since $\rho$ is a faithful representation
of $K_\rho$, which is compact, ${\cal H} (K_\rho)$ is generated by the
coefficients of $\rho$ and $\breve{\rho}$ [2] ; equivalently, $\theta$ is
a subrepresentation of a sum of tensor products of $\rho$ and $\breve{\rho}$.
Let us  denote this situation by $\theta \subset \hbox{Pol}_\otimes (\rho,
\breve{\rho})$. But $\gamma = \hbox{ Pol}_\otimes (\rho, \breve{\rho})$
is a representation of $G$, and $\theta \subset \gamma$ means that on some
subspace $W$, one has $\gamma \mid K\mid W = \theta \mid K$; therefore
$d\gamma (X) \mid_W = d \theta (X), X\in  {\goth g}$, and the integration
of $d \theta $ leads to a representation of $G$. So any representation of
$K_\rho$ is actually a representation of $G$. Now $K_\rho = U / {\cal Z}_\rho$,
where ${\cal Z}_\rho$ is a subgroup of the center of $U$, which is
finite, so when $\rho$ varies over all f.d. non trivial representations of $G$,
only a finite number of $K_\rho$ will appear. We denote them by $K_1,....,K_n$,
and by  $\rho_1,...., \rho_n$ the associated representations of $G$.

Given a non trivial representation $\rho$ of $G$, there exists $K_i$
such that $\rho$ is a faithful representation of $K_i$, and $\rho$,
as a representation of $K_i$, is contained in some Pol$_\otimes (\rho_i,
\breve{\rho}_i)$ [2] which is a representation of $G$.
Therefore the coefficients of $\rho$, as functions on $G$, are
polynomials in the coefficients of $\rho_i$ and $\breve{\rho}_i$.
It results that ${\cal H}(G)$ is generated, as an algebra, by the coefficients
of $\xi =  \rho_1\oplus ....\oplus \rho_n \oplus \breve{\rho}_1
\oplus....\oplus\breve{\rho}_n$. Also if
$N(G) = {\displaystyle{\bigcap_{\scriptstyle \rho\hskip.1cm
\hbox{\scriptsize f.d.}}}}\rho$,
one has $N(G) =\hskip.1cm $ker$\xi$, which
is the result of Harish Chandra quoted in (2.2). Now $\xi$ is also a
representation of $ K= U / \hskip.1cm \bigcap_i \hbox{Ker} \rho_i$,
and each $\rho_i$ is a  representation of $K$. Since $\xi$ is a faithful
self dual representation of $K$ which is compact, any representation
of $K$ is contained in some $\hbox{Pol}_\otimes (\xi)$, so any representation
of $K$ is actually a representation of $G$. On the other hand the
representations $\rho_i$ and $\breve{\rho}_i$, which generate by tensor
products all representations of $G$, are representations of $K$. This provides
a one to one mapping from Rep$(G)$ onto Rep$(K)$, preserving equivalence and
irreducibility, so $\Pi_G = \Pi_K =\Pi$, and then an isomorphism of Hopf
algebras ${\cal H} (G) \simeq {\cal H} (K)$ is defined by:
$\hbox{ If } \pi \in \Pi, M \in {\cal L} (V_\pi), \hbox{ define }
\phi_\pi (C_M^\pi \hbox{as\hskip.1cm a function \hskip.1cm on}\hskip.1cm G)
= (C_M^\pi\hskip.1cm \hbox{ as a function on } K), \hbox{ and then }
\phi = {\displaystyle \Sigma_\Pi} \phi_\pi.$
$\square$
\saut

{\bf{(2.6) Remarks}}: (1) In the case of a compact group $K$, any f.d. faithful
 and self dual representation $\pi$ is a complete set of representations of $K$
 (i.e. the coefficients of $\pi$ generate ${\cal H}(K)$), and such a
 representation always exist (see [2] and references quoted therein).
 For a noncompact semi-simple linear group $G$, from  the proof of (2.5),
 there exists a f.d. faithful and self dual representation $\pi$
which is a complete set, but is has to be noted that any f.d. faithful and
self dual representation needs not to be a complete set (see e.g. [9] p.116).
Nevertheless, by the proof of (2.2), the coefficients of any f.d. faithful
representation and of its conjugate always generate a dense subalgebra of
$C^\infty (G)$. So one has to be careful !
\saut

(2) (2.5) is a reformulation of the Weyl unitary trick. Equivalent formulations
using complex algebraic groups instead of compact Lie groups can be stated.

\gsaut

{\bf{3. Star-products on ${\cal H}(G)$ and $H(G)$.}}

We keep the notations and assumptions of Section 2.
In particular, $G$ is always assumed to
be a semi-simple connected Lie group. Let us consider a deformation ${\cal
U}_t\hskip.1cm \hbox{of } \hskip.1cm{\cal U} ={\cal U}({\goth g})$. As shown by
Drinfeld [4], it is automatically trivial, so we can assume ${\cal U}_t =
{\cal U}[[t]]$ with trivial product. Assume now that we have a coassociative
 coalgebra deformation of ${\cal U}_t$, with coproduct $\tilde{\Delta}$.
 By the argument of ([2] (6.2.1)), the initial counit $\varepsilon$
 is still a counit for the new structure, and by [7], there is an antipode,
 so we get finally a Hopf deformation. By [4], the new coproduct
 $\tilde{\Delta}$ is obtained from the initial one $\Delta_0$ by a twist,
 i.e. there exists $\tilde{P} \in {\cal U} \otimes {\cal U} [[t]]$ such
 that $\tilde{\Delta} = \tilde{P}
\Delta_o \tilde{P}^{-1}$. Set ${\cal A} = {\cal A}(G)$, and $A = A(G)$.
\saut

{\bf{(3.1) Lemma}}: {\it{$\hbox{ The formula } \tilde{\Delta} = \tilde{P}
 \Delta _0  \tilde{P}^{-1}$ defines a coassociative coproduct on
 ${\cal A}[[t]] \hbox{ and}  A[[t]]$.}}
\saut

{\bf{Proof}}: Since ${\cal U}$ is contained in $A$ and $\cal A$,
 $\tilde{\Delta}$ is clearly a morphism, and we have to show that
 it is still coassociative.

Let us start with a linear $G$. Then the proof is as in ([2](6.2.1)): one
has ${\cal U} \subset A \subset {\cal A}$ and $\overline{{\cal U}} = {\cal A}$,
and since the continuous maps ($\tilde{\Delta} \otimes I)\circ\tilde{\Delta}
\hbox{ and } (I \otimes \tilde{\Delta})\circ \tilde{\Delta}$ coincide on
$\cal U$, they coincide on $\cal A$, and a fortiori on $A$.
\psaut
Then we treat the case of general $G$. We introduce $K$ as in (2.5):
${\cal A}(G) = {\cal A}(K)$, and we are back to the linear case.
To get the result for $A$, we use the density of Vect$(G)$ in $A$ as follows.
Using the twist, our result will be proved if we prove the following:
if formula $(\Delta_0 \otimes I)\circ \Delta _0 = \tilde{\phi} \circ (I \otimes
\Delta_0) \circ \Delta_0 \tilde{\phi}^{-1}\hbox{, for some } \tilde{\phi}
\in {\cal U} \otimes {\cal U} \otimes {\cal U}[[t]]$, holds on ${\cal U}$,
then it holds on $A$. To prove that, we set $\pi_x = (\Delta_0 \otimes I)
\circ \hskip.1cm\Delta_0\hskip.1cm(x), \pi'_x  = \tilde{\phi}\circ (I \otimes
\Delta_0) \circ\Delta_0 (x)\circ \tilde{\phi}^{-1}$, and obtain two continuous
morphisms from $A$ into $A\hat\otimes A \hat\otimes A[[t]]$.
\psaut
Using the formula $\frac{d}{d\tau} (\hbox{exp} \tau X) = X \cdot
\hbox{exp}\tau X, X \in {\goth g}$, we have:
$$\frac{d}{d\tau} (\pi (\hbox{exp}\tau X)) = \pi (X) \pi (\hbox{exp} \tau X)
\hbox{ and } \frac{d}{d \tau} (\pi' (\hbox{exp}\tau X)) = \pi (X) \pi'(\hbox{
exp} \tau X).$$ Thus $\frac {d}{d \tau} (\pi ((\hbox{exp}\tau
X))^{-1} \pi'(\hbox{exp}\tau X)) =0$, so $\pi' (\hbox{exp}\tau X) =
\pi(\hbox{exp}\tau X)$.
But $G$ is connected, $\pi$ and $\pi'$ are morphisms, so we obtain that
$\pi'(x) = \pi (x), \forall x \in G$. From $\overline{\hbox{Vect}(G)} = A$,
we conclude that $\pi =  \pi'$.
$\square$
\saut

In order to have a Hopf structure on ${\cal A}[[t]]$, or $A[[t]]$, with
coproduct $\tilde{\Delta}$, we need a counit and an antipode.
For the counit, the initial one still works, and the proof is the same
as ([2] (4.2.6)). For the antipode, by [7], it does exist, and the next
problem is to show that it is an extension of the new antipode of
${\cal U}_t$. In the case of ${\cal A}[[t]]$, by (2.4) we can assume
that the group is compact, and then use ([2] (6.2.1)), so the result is true.

In order to prove the same for $A[[t]]$, we have to clarify the relations
between representations of $G$, $A(G)$ and ${\cal A}(G)$, as follows: any
given f.d. representation  $\pi$ of $G$, up to equivalence, splits into a
direct sum of representations in $\Pi$, which are all by definition
representations of ${\cal A}(G)$, so it extends to ${\cal A}(G)$.
Moreover, $\pi \in C^\infty (G, {\cal L}(V_\pi)) = C^\infty (G)
\hat\otimes {\cal L}(V_\pi) = {\cal L}(A(G), {\cal L}(V_\pi))$,
so $\pi$ extends to a continuous linear map (still denoted by $\pi$) from
$A(G) $ into ${\cal L} (V_\pi)$. It is well known that $\pi$ is a morphism
of algebras (cf. e.g. [14]), so it defines a representation of $A(G)$ in
$V_\pi$. Obviously, equivalence and irreducibility are preserved.
On the other hand the inclusion ${\cal H} (G) \subset H(G)$ provides,
by transposition, a continous linear morphism from
$A(G) $ into ${\cal A}(G)$, so any representation of ${\cal A}(G)$ is a
representation of $A(G)$; finally, any representation of $A(G)$ defines, by
restriction, a representation of $G$. So we have proved that the f.d.
representations of $G$, $A(G), $ or ${\cal A}(G)$ are the same.
\saut
\saut
{\bf{(3.2) Lemma}}:{\it{ The new antipode of $A[[t]]$ is an extension of
 the new  antipode of ${\cal U}[[t]]$.}}
\saut
{\bf{Proof}}: Let us denote by $S_{\cal U}, S_A $ and $S_{\cal A}$ the
 respective new antipodes of ${\cal U} [[t]], A [[t]]$ and ${\cal A} [[t]]$.
 Given a f.d. irreductible representation $\pi$, we consider $\tilde{\pi} =
{^{\scriptscriptstyle T}} \pi. S_A$ and by restriction to $\goth g$ we get a
 deformation  of the representation $\breve{\pi}$ of $\goth g$,
which is trivial, since $\goth g$ is semi-simple [10]. From this remark and
Burnside theorem we deduce that there exists $u_\pi \in {\cal U}[[t]]$ such
that :
$$\tilde{\pi} (u) = \breve{\pi} (u_\pi \cdot u \cdot u_\pi^{-1}),
\hskip.2cm \forall u \in {\cal U}.$$
We set $a=\displaystyle \Sigma_{ \pi \in \Pi} \hskip.1cm u_\pi \in
{\cal A} [[t]]$ and use the last formula to get :
$$\pi (S_A(u)) = \pi (S_0(a)^{-1} S_0 (u) S_0(a)), \forall u \in {\cal U},
\forall  \pi \in \Pi.$$
Let $S'= S_0(a)^{-1}\cdot S_0 \cdot S_0(a)$, denote by $m$ the product
of $A[[t]]$ or ${\cal A}[[t]]$, and write $\tilde{\Delta}(u) =
\displaystyle{\Sigma_i}\alpha_i \otimes \beta_i, u \in {\cal U}$, then :
$$\pi [m \circ( S' \otimes Id) \circ \tilde{\Delta}(u)] = \Sigma_i \pi (S_A
(\alpha_i)) \pi (\beta_i) = \pi [m \circ (S_A \otimes Id)
\circ \tilde{\Delta}(u)] = \pi [\varepsilon(u) 1],  $$
\noindent $\forall u \in {\cal U}, \pi \in \Pi.$
Since $\Pi$ separates points on $\cal U$ by (2.3), we get
 on $ {\cal U,}$  and by continuity on ${\cal A}$,
since  $\overline{\cal U} = {\cal A}:$
$$m \circ (S' \otimes Id ) \circ \tilde{\Delta} = \varepsilon.1$$
Therefore $S'$ is an antipode for ${\cal A}[[t]]$ with its new coproduct,
so using unicity of the antipode, $S' =  S_{\cal A}$ but since
$S_{\cal A}$ extends $S_{\cal U}$, we have $S'\mid{\cal U} = S_{\cal U}$,
and therefore:
$$\pi (S_A (u))= \pi (S_{\cal U}(u)), \forall u \in {\cal U},
\forall\pi \in \Pi.$$
Finally $S_A(u) -  S_{\cal A} (u) \in \displaystyle{\bigcap_{\pi \in
\Pi}}\hbox{Ker} \pi \bigcap {\cal U} = \{0\}$.
$\square$
\psaut
We can now state our main result :
\saut
{\bf{(3.3) Theorem:}}{\it{There exists a topological Hopf deformation of $A(G)$
(resp: ${\cal A}(G)$) which extends the Hopf deformation ${\cal U}_t \hbox{ or}
{\cal U}$.}}
\psaut
Now we apply the duality argument ([2] (3.8)):
\saut

{\bf{Corollary}}: {\it{The Hopf deformation ${\cal U}_t$ produces a preferred
 (i.e. unchanged coproduct and counit) topological Hopf deformation of
 ${\cal H}(G) $ and  $H(G)$.}}
\psaut

The new product on $H(G)$ and ${\cal H} (G)$ is the desired star product.
The above corollary applies for instance to the Drinfeld standard models [3],
and also to the Reshetikhin models [11]. It can be seen that the deformation of
${\cal H}(G)$ is the restriction of the deformation of $H(G)$, as in the
compact case [2]. In the linear case, if we start with the Drinfeld standard
model, then there exists a universal $R$-matrix satisfying Q.Y.B. equation [3],
and with the very proof given in [2], the star product on ${\cal H}(G)$
satisfies relations of the type  $R T_1  T_2 = T_2T_1 R$.

\gsaut
{\bf Acknowledgements.} We thank M. Flato and C. Fr\o nsdal for inspiring
 discussions and D. Sternheimer for helping in the final formulation of
  the paper.
\saut
\saut

\centerline{\bf REFERENCES.}
\gsaut

\noindent [1] F. Bayen, M. Flato, C. Fronsdal, A. Lichnerowicz and
D. Sternheimer: Deformation theory and quantization I and II,
Ann. Phys. {\bf{111}} (1978), 61 and 111.
\saut
\noindent [2] P. Bonneau, M. Flato, M. Gerstenhaber and G. Pinczon:
The hidden group structure of
quantum groups, Commun. Math. Phys., {\bf{161}}, (1994), 125.
\saut
\noindent [3] V.G. Drinfeld: Quantum Groups, in Proc. ICM 1986, AMS Providence,
(1987), {\bf{1}}, 798.
\saut

\noindent [4] V.G. Drinfeld: On almost cocommutative Hopf algebras,
Leningrad Math. J., {\bf{1}}, (1990), 321.
\saut

\noindent [5] L.D. Faddeev, N.Y. Reshetikhin and L.A. Takhtajan: Quantization
of Lie groups and
Lie algebras, Leningrad Math. J., {\bf{1}}, (1990), 193.
\saut

\noindent [6] M. Gerstenhaber: On the deformations of rings and algebras,
Ann. Math. {\bf{79}} (1964), 59.
\saut

\noindent [7] M. Gerstenhaber, S.D. Schack: Algebras, bialgebras, quantum
groups and algebraic deformations. Contemp. Math. {\bf{134}}, (1992), 51.
\saut

\noindent [8] Harish Chandra: Lie algebras and the Tannaka duality theorem,
Ann. Math. {\bf{51}}, (1950), 299.
\saut
\noindent [9] A.W. Knapp: Representation theory of semi-simple Lie groups.
Princeton Math. Series, {\bf{36}}, (1986).
\saut

\noindent [10] M. Lesimple, G. Pinczon: Deformations of representations of
Lie groups and Lie algebras, J. Math. Phys. {\bf{34}}, (1993), 4251.
\saut

\noindent [11] N. Reshetikhin: Multiparameter quantum groups,
Lett. Math. Phys. {\bf{20}}, (1990), {\bf{331}}.
\saut

\noindent [12] F. Tr\`eves: Topological vector spaces,
distributions and kernels, Academic Press (1967).
\saut

\noindent [13] E. Twietmeyer: Real forms of ${\cal U}_q({\goth g})$,
Lett. Math. Phys. {\bf 24} (1992), 49.
\saut

\noindent [14] G. Warner: Harmonic analysis on semi-simple Lie groups I,
Springer Verlag, (1972).
\saut

\noindent [15] S.L. Woronowicz: Compact matrix pseudogroups.
Commun. Math. Phys. {\bf{111}}, (1987), 613.
\saut
\end{document}